\documentclass{iaus}
\usepackage{graphicx}
\AtBeginDvi{}

\title[Type Ia Supernova Models] %% give here short title %%
{Type Ia Supernova Models and \\ Progenitor Scenarios}

\author[K. Nomoto \etal]   %% give here short author list %%
{Ken'ichi Nomoto$^{1,\dagger}$, Yasuomi Kamiya$^{2,1}$, \and Naohito Nakasato$^{3}$
}

\affiliation{
$^1$Kavli Institute for the Physics and Mathematics of the Universe, The University of Tokyo,\\
5-1-5 Kashiwanoha, Kashiwa, Chiba 277-8583, Japan\\[\affilskip]
$^2$Department of Astronomy, Graduate School of Science, The University of Tokyo,\\
7-3-1 Hongo, Bunkyo-ku, Tokyo 113-0033, Japan\\[\affilskip]
$^3${Department of Computer Science and Engineering, University of Aizu,\\
Aizu-Wakamatsu, Fukushima 965-8580, Japan}\\[\affilskip]
$^\dagger$email: {\tt nomoto@astron.s.u-tokyo.ac.jp}
}

\pubyear{2011}
\volume{281}  %% insert here IAU Symposium No.
\pagerange{}
\jname{Binary Paths to Type Ia Supernova Explosions}
\editors{R. Di Stefano, M. Orio, \& M. More, eds.}

\renewcommand{\(}{\left(}
\renewcommand{\)}{\right)}
\newcommand{\ms}{$M_{\odot}$}
\newcommand{\Ms}{M_\odot}
\newcommand{\lsim}{\mathrel{\rlap{\lower 4pt \hbox{\hskip 1pt $\sim$}}\raise 1pt \hbox
        {$<$}}}
\newcommand{\gsim}{\mathrel{\rlap{\lower 4pt \hbox{\hskip 1pt $\sim$}}\raise 1pt \hbox
        {$>$}}}
\newcommand{\nat}{\textit{Nature}}

\newcommand{\apj}{\textit{ApJ}}
\newcommand{\apjs}{\textit{ApJS}}
\newcommand{\araa}{\textit{ARAA}}
\newcommand{\mnras}{\textit{MNRAS}}

\newcommand{\aap}{\textit{A\&A}}
\newcommand{\pasj}{\textit{PASJ}}
\newcommand{\aipcp}{\textit{AIPC}}
\newcommand{\sci}{\textit{Science}}
\newcommand{\MWD}{M_\mathrm{WD}}
\newcommand{\Menv}{M_\mathrm{env}}

\begin{document}

\maketitle

\begin{abstract}

We review some recent developments in theoretical studies on the
connection between the progenitor systems of Type Ia supernovae (SNe
Ia) and the explosion mechanisms.  (1) {\sl DD-subCh}: In the merging
of double C+O white dwarfs (DD scenario), if the carbon detonation is
induced near the white dwarf (WD) surface in the early dynamical
phase, it could result in the (effectively) sub-Chandrasekhar mass
explosion.  (2) {\sl DD-Ch}: If no surface C-detonation is ignited,
the WD could grow until the Chandrasekhar mass is reached, but the
outcome depends on whether the quiescent carbon shell burning is
ignited and burns C+O into O+Ne+Mg.  (3) {\sl SD-subCh}: In the single
degenerate (SD) scenario, if the He shell-flashes grow strong to
induce a He detonation, it leads to the sub-Chandra explosion.  (4)
{\sl SD-Ch}: If the He-shell flashes are not strong enough, they still
produce interesting amount of Si and S near the surface of C+O WD
before the explosion.  In the Chandra mass explosion, the central
density is high enough to produce electron capture elements, e.g.,
stable $^{58}$Ni.  Observations of the emission lines of Ni in the
nebular spectra provides useful diagnostics of the sub-Chandra
vs. Chandra issue.  The recent observations of relatively low velocity
carbon near the surface of SNe Ia provide also interesting constraint
on the explosion models.

\end{abstract}

\firstsection % if your document starts with a section,
\section{Introduction}

The observed features of Type Ia supernovae (SNe Ia) have been
well-understood by a thermonuclear explosion of a carbon-oxygen (C+O)
white dwarf (WD).  Both the Chandrasekhar mass [{\sl Chandra (Ch)}
  model] and the sub-Chandrasekhar mass [{\sl sub-Chandra (subCh)}
  model] have been presented (e.g., \cite{liv00}).  However, there has
been no clear observational indication as to how the WD mass grows
until carbon ignition; i.e., whether the WD accretes H/He-rich matter
from its binary companion [{\sl single degenerate (SD)} scenario], or
two C+O WDs merge [{\sl double degenerate (DD)} scenario] (e.g.,
\cite{hil00,nom97}, 2000, 2009, \cite{arn96}).  Even before these
issues are resolved, several candidates of {\sl super-Chandrasekhar}
mass explosions have been observed (e.g., \cite{hachi12} and
references therein).

Recent modeling shows that DD merging could result in both the Chandra
and the (effectively) sub-Chandra explosions.  The SD scenario could
also result in both the Chandra and sub-Chandra explosions.
Here we review such cross connections between (DD, SD) scenarios and
(Chandra, sub-Chandra) models, and some observational constraints.

(1) {\sl DD-subCh}: In the DD scenario, if the carbon detonation is
induced near the WD surface in the early dynamical phase, it could
result in the (effectively) sub-Chandra explosion.  (2) {\sl DD-Ch}:
If no detonation is induced, the WD could grow until the Chandrasekhar
mass is reached.  The outcome depends on whether the quiescent carbon
shell-burning is ignited and burns interior C+O into O+Ne+Mg.  (3)
{\sl SD-subCh}: In the SD scenario, if the He shell-flashes grow
strong to induce a He detonation, it leads to the sub-Chandra
explosion.  (4) {\sl SD-Ch}: If the He-shell flashes are not strong
enough to induce a He detonation, such flashes produce interesting
amount of intermediate mass elements, including Si and S, as 
unburned material near the surface of C+O WD.

\section{C+O Double Degenerates to Sub-Chandra Mass Explosion} %%%%%%%%%%%%%%%

In the DD scenario (\cite{ibe84,web84}), two C+O WDs form a close
binary system after the common envelope phase and get closer by losing
orbital angular momentum due to gravitational radiation.  Eventually,
the less massive WD with a mass $M_2$ fills its Roche lobe and
dynamically evolves into the formation of a massive disk/envelope
around the more massive WD with a mass $M_1$.

Such a dynamical evolution of WD binary has been studied by number of
authors (\cite{ben90,seg97,gue04,shi07,pak10}) using the Smoothed
Particle Hydrodynamics (SPH) method.

{\bf Surface Carbon Detonation:} The important question is whether the
merging process ignites a surface carbon detonation in the WD.  During
the early merging process, the shock heating increases the temperature
in the surface C+O layer of the primary (i.e., more massive) WD.  If
the temperature becomes high enough to induce C-detonation
(\cite{pak10} adopted the critical temperature of 2.5 $\times 10^9$
K), the detonation wave propagates through the central region of the
primary WD.  Eventually, the whole primary WD is detonated and
disrupted.

This model is essentially the sub-Chandra mass explosion, because
the detonated WD has a sub-Chandra mass of $M_1 = 0.9$--1.1\ms\ and
the central density is as low as $\sim 10^7$ g cm$^{-3}$
(\cite{pak10}, 2011).  The explosion produce a larger amount of
$^{56}$Ni for larger $M_1$.

\section{Double Degenerates to Chandra Mass Explosion or Collapse} %%%%%%%%%%%%%%%

Suppose the surface C-detonation is not triggered in the early
dynamical phase of merging.  Then the next important question is
whether the merging process ignites not the C-detonation but
``quiescent'' off-center carbon burning in the WD.  Once carbon is
ignited, carbon flame moves inward by heat conduction to reach the
center (\cite{sai85}, 1998).  The released nuclear energy is
lost in neutrinos, and the C+O WD is converted into the O+Ne+Mg WD
non-explosively.  The ONeMg WDs eventually collapse due to electron
capture rather than exploding as SNe Ia (\cite{nom84}, 1987).

\subsection{Carbon Shell Burning and Chandra Explosion or Collapse} %%%%%%%%%%%%%%%

% Fig1: tmax

\begin{figure}
\begin{center}
  \includegraphics[clip,angle=270,totalheight=.328\textheight]{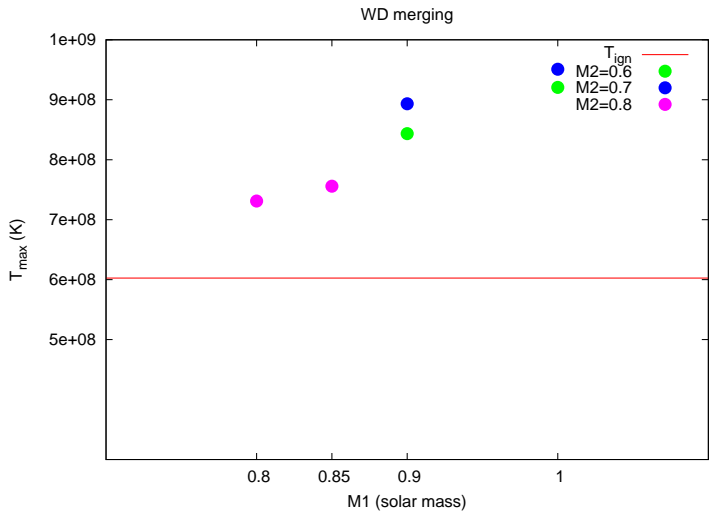}
\caption{The local peak temperature $T_{\rm peak}$ attained 2.5 min
after the start of merging of double C+O white dwarfs with masses ($M_1$,
$M_2$) (\cite{nak11}).  For $T_{\rm peak} > T_{\rm ign}$, off-center
carbon burning is ignited.}
\label{tmax}
\end{center}
\end{figure}
%%%%%%%%%%%%%%%%%%%%%%%%%%%%%%%%%%%%%%%%%%%%%%%%%%%%

Here the carbon ignition temperature $T_{\rm ign}$ is defined by
$\epsilon_{\rm C+C} = \epsilon_{\nu}$, where $\epsilon_{\rm C+C}$ and
$\epsilon_{\nu}$ denote the nuclear energy generation rate and the
neutrino energy loss rate, respectively.  At $\rho \sim 1$--$3 \times 10^6$ g
cm$^{-3}$, $T_{\rm ign} \sim 6 \times 10^8$ K (\cite{nom85,kawa87,yoon07}).

For $T > T_{\rm ign}$, $\epsilon_{\rm C+C} > \epsilon_{\nu}$ and the
carbon flash is ignited, and the conductive carbon flame propagates
inward.  If $T < T_{\rm ign}$, on the contrary, $\epsilon_{\rm C+C} <
\epsilon_{\nu}$ and the neutrino cooling dominates to induce the
gradual contraction of the C-rich envelope.  Then carbon flame is not
formed.

There are two possible cases for the off-center carbon ignition to
occur.

\noindent
{\bf Case 1}: When the falling materials from the less massive star
compress the outer layer of the more massive star, the material could
be heated up to high enough temperature to ignite carbon burning.

\noindent
{\bf Case 2}: Later, if the accretion rate exceeds a critical rate of
$2.7 \times 10^{-6} M_\odot$ yr$^{-1}$, compressional heating exceeds radiative
cooling and leads to carbon ignition (\cite{nom85,kawa87}).

For Case 1, \cite[Yoon \etal\ (2007)]{yoon07} calculated the merging process
until 5 min after its start and showed that a quasi-static equilibrium
configuration is reached consisting of a cold core, hot envelope, and
a disk.  The peak temperature $T_{\rm peak}$ reaches a stationary
value.  For $M_1 = 0.9$\ms\ and $M_2 = 0.6$\ms, $T_{\rm peak}$ is
marginally lower than $T_{\rm ign}$.  Whether the later off-center
carbon ignition of Case 2 takes place depends on the effective
accretion rate, which needs further study (e.g., \cite{shi07,shen12}).

For Case 1, \cite[Nakasato \& Nomoto (2011)]{nak11} have recently conducted
the SPH simulation (\cite{nak03}) with the number of particles $N =
300,000$ and $N = 1,000,000$ for various combinations of ($M_1$,
$M_2$).  The artificial viscosity is treated to minimize numerical
effects according to the hybrid scheme proposed by
\cite[Rosswog \etal\ (2010)]{ros00}.  The Helmholtz equation of state is used
(\cite{tim00}), and the WD is assumed to be composed of 50 \% of carbon
and 50 \% of oxygen.

The local peak temperatures shown in Figure \ref{tmax} are obtained
from particles at $\rho \sim 1$--$3 \times 10^6$ g cm$^{-3}$ at 2.5 min after
the start of merging.  $T_{\rm peak}$ has already reached its
stationary value, which is confirmed with some test runs calculated
until 10 min after the merging.  It is seen that $T_{\rm peak}$ is
determined mainly by $M_1$ with small dependence on $M_2$.  This
implies that the gravitational potential of the more massive WD is the
main factor to determine $T_{\rm peak}$.

Figure \ref{tmax} shows that $T_{\rm peak} > T_{\rm ign}$ for most
cases of ($M_1$, $M_2$).  Thus carbon flash will take place to form
a carbon flame that propagates inward to convert C+O into O+Ne+Mg.

How the above results depend on the set-up of the merging calculations
(\cite{dan11})?  The initial separation is set to be $a = (0.9
R_2)/r_{\rm L}$ where $R_2$ is the radius of the less massive WD and
$r_{\rm L}$ is the effective Roche lobe radius.  Some test runs
adopting a larger separation confirm that $T_{\rm peak}$ is not
sensitive to the initial separation.

Comparisons between the runs with $N = 300,000$ and with $N =
1,000,000$ show that $T_{\rm peak}$ of the higher resolution run is
always slightly higher than the low resolution run.  Thus carbon
ignition of Case 1 is quite likely for most cases of ($M_1, M_2$).

\subsection{Further Evolution of Rotating White Dwarfs} %%%%%%%%%%%%%%%

The WD formed from merging must be rapidly rotating.  The ``SN Ia
mass'' of the WD with which SN Ia is triggered is $M_{\rm Ia} =
1.48$\ms\ for a uniformly rotating WD at the critical rotation
(\cite{ueni03}).  This is larger than 1.38\ms\ for the non-rotating
WD.  For non-uniform, differentially rotating WDs, $M_{\rm Ia}$ is as
large as $\gsim 2$\ms\ (\cite{yoon04,hachi12}).  Therefore, even if no
carbon flame is formed, the WD may not reach $M_{\rm Ia}$ because of
some mass loss after merging.

\section{Single Degenerate to Sub-Chandra Explosion}

In the SD scenario, H-burning produces a thin He layer, and He-flashes
are ignited when the He mass reaches a certain critical value.  The
strength depends on the He envelope mass $\Menv$, thus depending on the
accretion rate.

The He envelope mass $\Menv$ is larger for the slower
mass-accumulation rate of the He layer $\dot{M_{\rm He}}$.  For
$\dot{M_{\rm He}} \gsim 1 \times 10^{-8}$ \ms\ yr$^{-1}$, $\Menv$
exceeds a critical value where the density at its bottom becomes high
enough to induce a He detonation.  This would result in the
sub-Chandra explosion (e.g., S. Sim in this conference).

%%%%%

\section{Single Degenerate to Chandra Mass Explosion}

% Fig. 2

\begin{figure}[t]%%%%%%%%%%%%%%%
  \begin{center}
   \includegraphics[clip,height=.328\textheight]{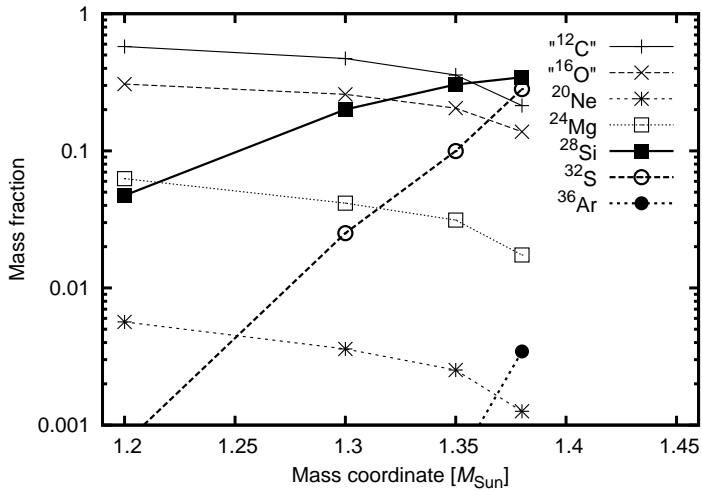}
  \end{center}
%  \vspace{-1pc}
  \caption{Abundance distribution of the products of He shell flashes
in an accreting WD just before the explosion (Kamiya \& Nomoto 2011).}
\label{he_flash}
\end{figure}%%%%%%%%%%%%%%%

For higher $\dot{M}_{\rm He}$, the He-shell flashes are not strong
enough to induce a He detonation.  Then, such flashes repeat many
times with the increasing WD mass $\MWD$ toward the Chandrasekhar
mass.  \cite{kami11} calculated nucleosynthesis in such He
shell flashes for various set of ($\MWD$, $\Menv$) (see also
\cite{shen07}).

For $\dot{M}_{\rm He}\sim (0.5 - 1) \times 10^{-7}$ \ms\ yr$^{-1}$, 
the WD is expected to undergo He shell-flashes
at $(\MWD/\Ms,\log(\Menv/\Ms))\sim \(1.2,-2.5\)$, $\(1.3,-3\)$,
$\(1.35,-3.5\)$, $\(1.38,-4\)$ as $\MWD$ grows (\cite{kato08}).

In the early stages of the He shell-flash, the envelope is
electron-degenerate and geometrically almost flat.  Thus the
temperature at the bottom of the He-burning shell increases because of
the almost constant pressure there.  Heated by nuclear burning, the
helium envelope gradually expands, which decreases the pressure.
Then, the temperature attains its maximum and starts decreasing.  The
maximum temperature is higher for more massive WD and more massive
envelope because of higher pressure.

For higher maximum temperatures, heavier elements, such as $^{28}$Si
and $^{32}$S, are synthesized.  However, the maximum temperature is
not high enough to produce $^{40}$Ca.  After the peak, some amount of
He remains unburned in the flash and burns into C+O during the stable
He shell burning.

In this way, it is possible that interesting amount of intermediate
mass elements, including Si and S, already exist in the unburned C+O
layer at $M_r\geq1.2$\ms.  Such a distribution is shown in
Figure~\ref{he_flash} for $\dot{M}_{\rm He}\sim 5 \times 10^{-8}$
\ms\ yr$^{-1}$.  A part of them might be ejected out.  Therefore the
above quantity of the synthesized elements are overestimated.

\section{Electron Capture in Chandra Mass Models} %%%%%%%%%%%%%%%

% Fig. 3

\begin{figure}[t]%%%%%%%%%%%%%%%
  \begin{center}
    \includegraphics[width=.58\textwidth]{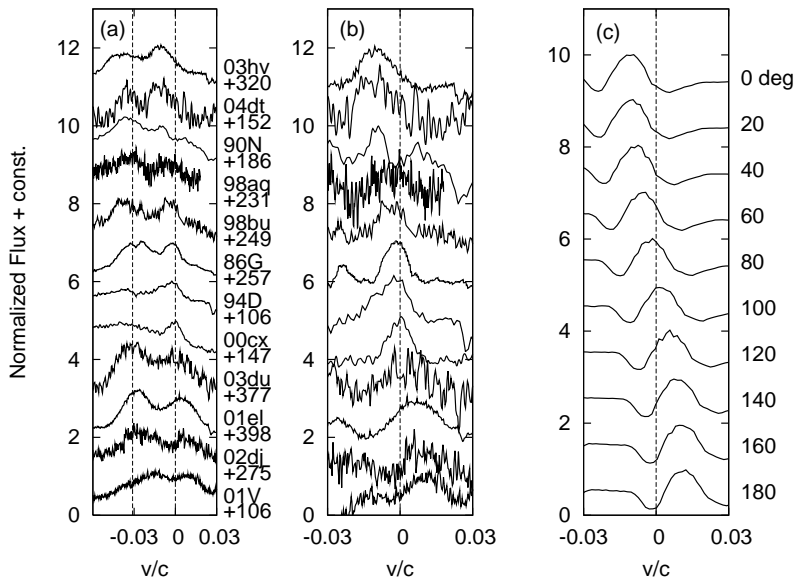}
  \end{center}
%  \vspace{-1pc}
  \caption{Analysis of the [Ni~II]~$\lambda$7378 line profiles in 12
SNe~Ia (\cite{maeda10a}).  The velocity is set assuming that the
rest wavelength is at 7378~\AA.
(a) Observed line profiles. The rest wavelengths of
[Fe~II]~$\lambda$7155 and [Ni~II]~$\lambda$7378 are shown by dotted
lines.
(b) [Ni~II]~$\lambda$7378 in observations, after removing the
underlying continuum (or possible other lines).
(c) Synthetic line profiles of the [Ni~II], depending on the viewing
orientation.}
\label{ni58}
\end{figure}%%%%%%%%%%%%%%%

%%%%%

Both Chandra and sub-Chandra explosion models can synthesize relevant
amount of $^{56}$Ni for SNe Ia.  However, the amount of other Fe-peak
elements differs, because the ignition density is different, being as
high as $> 10^9$ g cm$^{-3}$ in the Chandra model, while as low as
$\sim 10^7$ g cm$^{-3}$ in the sub-Chandra model.

In the Chandra model, the thermonuclear runaway starts with the
ignition of deflagration (e.g., \cite{nom76}, 1984).  In the
high temperature and density bubble, materials are incinerated into
NSE (nuclear statistical equilibrium) and undergo electron capture.
Electron capture on free protons and Fe-peak elements leads to the
synthesis of $^{58}$Ni, $^{54}$Fe, and $^{56}$Fe (not via $^{56}$Ni
decay).  These neutron-rich Fe-peak elements form an almost
$^{56}$Ni-empty hole (e.g., \cite{nom84b}).

In the sub-Chandra model, the ignition density is too low for electron
capture to take place.  The neutron excess is produced only by the
initial CNO elements which are converted to $^{14}$N and to $^{22}$Ne,
thus depending on the initial metallicity.  As a result, the mass
fraction of $^{58}$Ni is as small as $\sim$ 0.01 (e.g.,
\cite{shige92}).

Such a difference in the amount of $^{58}$Ni can be observationally
investigated by late-phase ($\sim 1$yr since the explosion)
spectroscopy at near-infrared (NIR) wavelength.  Because the ejecta
become optically thin in late phases, spectroscopy provides an
unbiased, direct view of the innermost regions.

Figure~\ref{ni58} shows the spectral feature around $\sim 7000 -
7500$~\AA, i.e., [Fe~II]~$\lambda7155$ (with some contribution from
[Fe~II]~$\lambda7171$) and [Ni~II]~$\lambda$7378, for 12 SNe~Ia
(\cite{maeda10a}).  The [Ni~II]~$\lambda$7378 line is emitted from the
electron capture region of the ejecta, which is supported by the
relatively narrow width ($\lsim 3,000$~km~s$^{-1}$) of the [Ni~II]
line.  Thus the existence of [Ni~II] line implies the ignition at
high density, thus favoring the Chandra model.

It is also interesting to note that these emission lines show the
velocity shift, which indicates the off-center ignition and aspherical
nature of SN Ia explosions (\cite{maeda10a}, 2010b).  The aspherical
features could be seen in the light echo from Tycho's supernova
remnant (\cite{usuda11}).

\section{Carbon in SNe Ia} %%%%%%%%%%%%%%%

% Fig. 4

\begin{figure}[t]%%%%%%%%%%%%%%%
  \begin{center}
   \includegraphics[clip,angle=270,totalheight=.328\textheight]{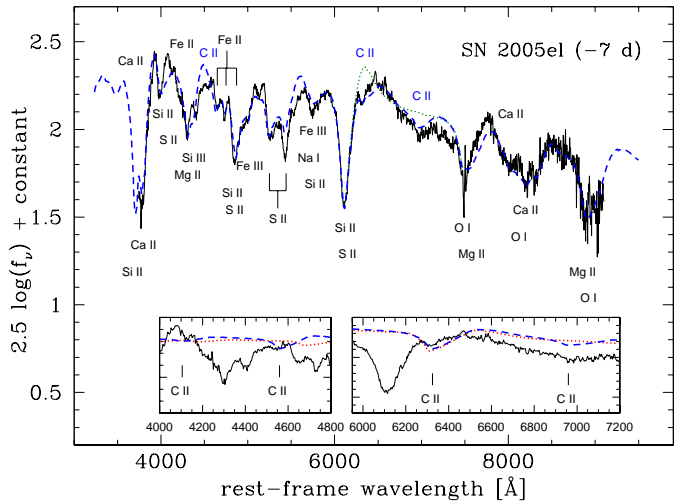}
  \end{center}
%  \vspace{-1pc}
  \caption{Spectrum of SN~2005el at $-7$ days ({\em solid line}), and
    a matching SYNOW calculation ({\em dashed line}). ({\em Insets})
    The spectrum near the location of C~II lines ({\em solid line}),
    the SYNOW spectra containing only C~II ({\em dashed line}), and
    only H~I ({\em dotted line}).  The expected locations of C~II lines
    are marked (Folatelli et al. 2012)}
\label{carbon}
\end{figure}%%%%%%%%%%%%%%%

\cite{parrent11} and \cite{fola12} investigated the presence of
unburned material in early-time spectra of SNe Ia.  They find that at
least 30\% of the objects in the sample show absorption at about
6300 \AA\ which can be associated with C~II $\lambda$6580
(Figure~\ref{carbon}). This would imply a larger incidence of carbon in
SN Ia ejecta than previously noted.  If confirmed as carbon, the material
producing the observed features would be present at very low expansion
velocities, merely $\sim$1,000 km s$^{-1}$ above the Si~II velocities.

Carbon must be present in very deep regions, corresponding to
velocities as low as $v$ $\approx$ 11,000 km s$^{-1}$. This is well
below the expected limit imposed by one-dimensional models, and points
directly to large mixing effects and/or possible departures from
spherical symmetry or clumpiness.

In spherically symmetric models, irrespective of the details of the
flame propagation (deflagration or detonation), the production of a
large amount of $^{56}$Ni ($\sim 0.6 M_{\odot}$) requires that the
strength of the flame, which must lead to the total consumption of
carbon below 15,000 - 20,000 km s$^{-1}$ (\cite{nom84b,shige92,iwa99}).
The situation is different for non-spherical models.  For example, the
off-center ignition model (\cite{maeda10a}; see also \cite{kasen09})
synthesizes $0.54 M_{\odot}$ of $^{56}$Ni and carbon with a mass
fraction of $\sim 0.1$ at velocities as low as $\sim$13,000 km
s$^{-1}$.  This is still larger than the observed velocity, but
suggests that the non-spherical effects may be important to understand
the detection of carbon deep in the ejecta.

The existence of carbon at relatively low velocity suggests the
presence of fair amount of unburned C+O materials.  The amount Fe peak
elements in such an unburned layer depends on the metallicity.
Further observations of early UV spectra could show significant
metallicity effects.

\section{Concluding Remarks} %%%%%%%%%%%%%%%

As summarized above, recent modeling shows that DD merging could
result in both the Chandra and the sub-Chandra explosions depending on
whether a carbon detonation is induced near the surface of more
massive WD.  The SD scenario could also result in both the Chandra and
sub-Chandra explosions depending on whether the He shell-flashes near
the surface of the WD induce a He detonation.

\begin{itemize}

\item For DD-subCh, it is critically important to confirm the
  formation of surface carbon detonation by means of 3D
  hydrodynamical simulations rather than SPH method.

\item For DD-Ch, whether carbon ignition can occur for both Case 1 and
  Case 2 needs further study for various set of ($M_1, M_2$).

\item For SD-subCh, a mechanism to avoid the production of too much
  $^{56}$Ni in the surface should be studied.  In case of the He-WD
  and C-WD merger, formation of an extended He envelope needs to be
  avoided.

\item For SD-Ch, the outcome of quiescent He-shell flashes, e.g.,
  nucleosynthesis, the rate of the He wind mass loss, needs to be
  studied.

\item Finally, detailed hydrodynamical modeling of the WD spin-down is
necessary.  In this spin-down scenario (e.g.,
\cite{just11,stefano11,soker12,hachi12}), an almost uniformly rotating
C+O WD with a mass range of 1.38 - 1.48 \ms\ forms and eventually
contract to ignite carbon after a long spin-down time.  The outcome
depends on the ignition density.  If it is as high as $\sim 10^{10}$ g
cm$^{-3}$, electron capture induces collapse rather than explosion
(\cite{nom91}).  If it is lower, an SN Ia explosion would
result.

\end{itemize}

\bigskip

This research has been supported in part by World Premier
International Research Center Initiative, MEXT, Japan, and by the
Grant-in-Aid for Scientific Research of the JSPS (23540262) and MEXT
(22012003, 23105705).

\end{document}